\documentclass[showpacs,amsmath,amssymb,superscriptaddress]{revtex4-1}

\usepackage{graphicx}
\usepackage{dcolumn}
\usepackage{bm}

\begin{document}

\title{Microwave-dressed state-selective potentials for atom interferometry}
\author{V. Guarrera}
\email{vera.guarrera@obspm.fr}
\affiliation{LNE-SYRTE Observatoire de Paris, CNRS, UPMC, 75014 Paris, France}
\author{R. Szmuk}
\affiliation{LNE-SYRTE Observatoire de Paris, CNRS, UPMC, 75014 Paris, France}
\author{J. Reichel}
\affiliation{Laboratoire Kastler Brossel, ENS, UPMC-Paris 6, CNRS, College de France,
24 rue Lhomond, 75005 Paris, France}
\author{P. Rosenbusch}
\email{peter.rosenbusch@obspm.fr}
\affiliation{LNE-SYRTE Observatoire de Paris, CNRS, UPMC, 75014 Paris, France}

\date{\today}

\begin{abstract}
We propose a novel and robust technique to realize a beam splitter for trapped Bose-Einstein condensates (BECs). The scheme relies on the possibility of producing different potentials simultaneously for two internal atomic states. The atoms are coherently transferred, via a Rabi coupling between the two long-lived internal states, from a single well potential to a double-well. We present numerical simulations supporting our proposal and confirming excellent efficiency and fidelity of the transfer process with realistic numbers for a BEC of $^{87}$Rb. We discuss the experimental implementation by suggesting state-selective microwave potentials as an ideal tool to be exploited for magnetically trapped atoms. The working principles of this technique are tested on our atom chip device which features an integrated coplanar micro-wave guide. In particular, the first realization of a double-well potential by using a microwave dressing field is reported. Experimental results are presented together with numerical simulations, showing good agreement. Simultaneous and independent control on the external potentials is also demonstrated in the two Rubidium clock states. The transfer between the two states, featuring respectively a single and a double-well, is characterized and it is used to measure the energy spectrum of the atoms in the double-well. Our results show that the spatial overlap between the two states is crucial to ensure the functioning of the beamsplitter. Even though this condition could not be achieve in our current setup, the proposed technique can be realized with current state-of-the-art devices being particularly well suited for atom chip experiments. We anticipate applications in quantum enhanced interferometry.        
\end{abstract}

\pacs{03.75.Hh, 03.75.Kk, 05.30.Jp}

\maketitle

\tableofcontents


\section{INTRODUCTION}
In recent years, Bose-Einstein condensates (BECs) have proven to be appealing systems for the realization of atom interferometers, owing to their properties of macroscopic phase coherence \cite{cronin2009}. Furthermore the possibility to interact with the environment, including electro-magnetic radiation, and their inertial nature promote them as powerful sensors. Interparticle interactions are also a fundamental ingredient as they are responsible for multi-particle entanglement which is known to lead to a phase estimation uncertainty below the standard shot-noise limit. Recently, few experiments with BECs have initiated this path by exploiting particle squeezing \cite{esteve2008,gross2010,reidel2010}. Even though they are not yet better performing than the best state-of-the-art atom interferometers, which are mainly limited by shot-noise (the standard quantum limit), BEC based in-trap interferometers will be crucial for local probing on the scale of few microns, due to their small size, and for application requiring a relatively low atom number. Possible usages include compact electro-magnetic and inertial forces sensing devices. 
State-of-the-art BECs interferometers suffer from technical noise as finite temperature effects, excitations, detection noise as well as phase diffusion in configurations working with trapped atomic ensembles \cite{lewenstein1996} which limit the phase sensitivity above the standard quantum limit. Reducing technical noise in the process of beam splitting and read out as well as elaborating new strategies that can lead to reach and also beat the shot-noise limit are nowadays crucial topics. In particular, commonly employed techniques to split the atomic cloud in a trap can cause excitations limiting the coherence time \cite{shin2004, schumm2005} and involve dynamics affecting the initial phase coherence between the two separated clouds in a non-trivial way \cite{berrada2013}. Much effort has been dedicated through the years to optimize these same techniques while novel ones, capable of definitely solving these problems, have not been found yet.    
 
To accomplish this task, we propose here a novel scheme for a robust and fast beam splitter with trapped Bose-Einstein condensates which does not require a dynamical variation of the trapping potential, neither a momentum transfer to the atoms. The atoms are initially prepared in an internal state $\vert 1 \rangle$ in a single well potential. At the same time a double-well potential is generated selectively for a second internal state $\vert 2 \rangle$. The splitting is realized by transferring the atoms from state $\vert 1 \rangle$ to state $\vert 2 \rangle$ by means of a $\pi$-pulse of coherent electromagnetic radiation. We perform numerical simulations of this scheme, demonstrating nearly unitary efficiency in the transfer to the ground state of the double-well potential. As it concerns the experimental implementation, this requires the realization of external potentials which are different for the two internal states, $\vert 1 \rangle$ and $\vert 2 \rangle$, and to create a double-well for the state $\vert 2 \rangle$. Few techniques exist to drive state-dependent potentials: radio-frequency dressing or micro-wave dressing of magnetically trapped states and dipole forces by optical beams with proper wavelength and polarization. The solution which we consider here implies the use of state-selective microwave-dressed potentials. We report experimental results showing that the proposed splitter can be notably realized on a compact atom chip device, by a micro-wave near-field. 

This paper is organized as follows: in section II we discuss the proposed scheme in a detailed way, analysing its features and constraints. We present numerical simulations based on coupled Gross-Pitaevskii equations, demonstrating the feasibility of the proposal under realistic experimental conditions. We consider, in particular, Bose-Einstein condensates of $^{87}$Rb in their two hyperfine ground states.  
In section III we investigate the use of microwave dressing to shape the underlying magnetic trap and to obtain independently controlled external potentials for the two different internal states. We exploit our atom chip device to experimentally prove the working principles of this strategy. We report, in particular, on the first realization and characterisation of a microwave-dressed double-well potential. In section IV we finally prove independent and simultaneous control over the external potentials for the two hyperfine states $\vert F=1, m_F=-1\rangle$ and $\vert F=2, m_F=1\rangle$ of $^{87}$Rb on our atom chip. We characterize the atom transfer between these two levels and we demonstrate the importance of independently dressing the two states in order to adjust the spatial overlap of the clouds to optimize the process. A measurement of the atomic energy spectrum in the double well potential can be derived for different intensities of the dressing field. Finally we discuss further experimental developments by means of state-of-the-art devices and possible applications of the presented technique.
    
\section{A BEAM SPLITTER BASED ON STATE-SELECTIVE POTENTIALS}

A common way to split a BEC for interferometric purposes consists of dynamically raising a potential barrier in the centre of the atomic distribution. The potential in which the atoms sit thus gradually evolves from a harmonic one to a double-well. For a given barrier height, the properties of the final state strongly depend on the duration of the splitting ramp and on the interatomic interactions \cite{menotti2001,streltsov2007,faust2010}. Long ramping-up times are generally required to reduce unwanted excitations, making the process as adiabatic as possible. Note that during the ramping-up the energy of the first excited state progressively approaches the one of the ground state to the point of becoming degenerate. An ideally adiabatic transfer cannot thus be reached even for long times, i.e. longer than the typical BEC lifetimes of standard experiments \cite{streltsov2007}, and excitations are produced. As a consequence subsequent manipulations and interrogation of the atomic systems should be performed faster than the timescales of the detrimental processes initialized during the splitting. On the other hand, the preservation of the phase coherence between the two separated clouds imposes some stringent limitation on the maximum duration of the ramping-up dynamics \cite{faust2010}. 
\begin{figure}
\centering
\includegraphics[width=0.2\textwidth]{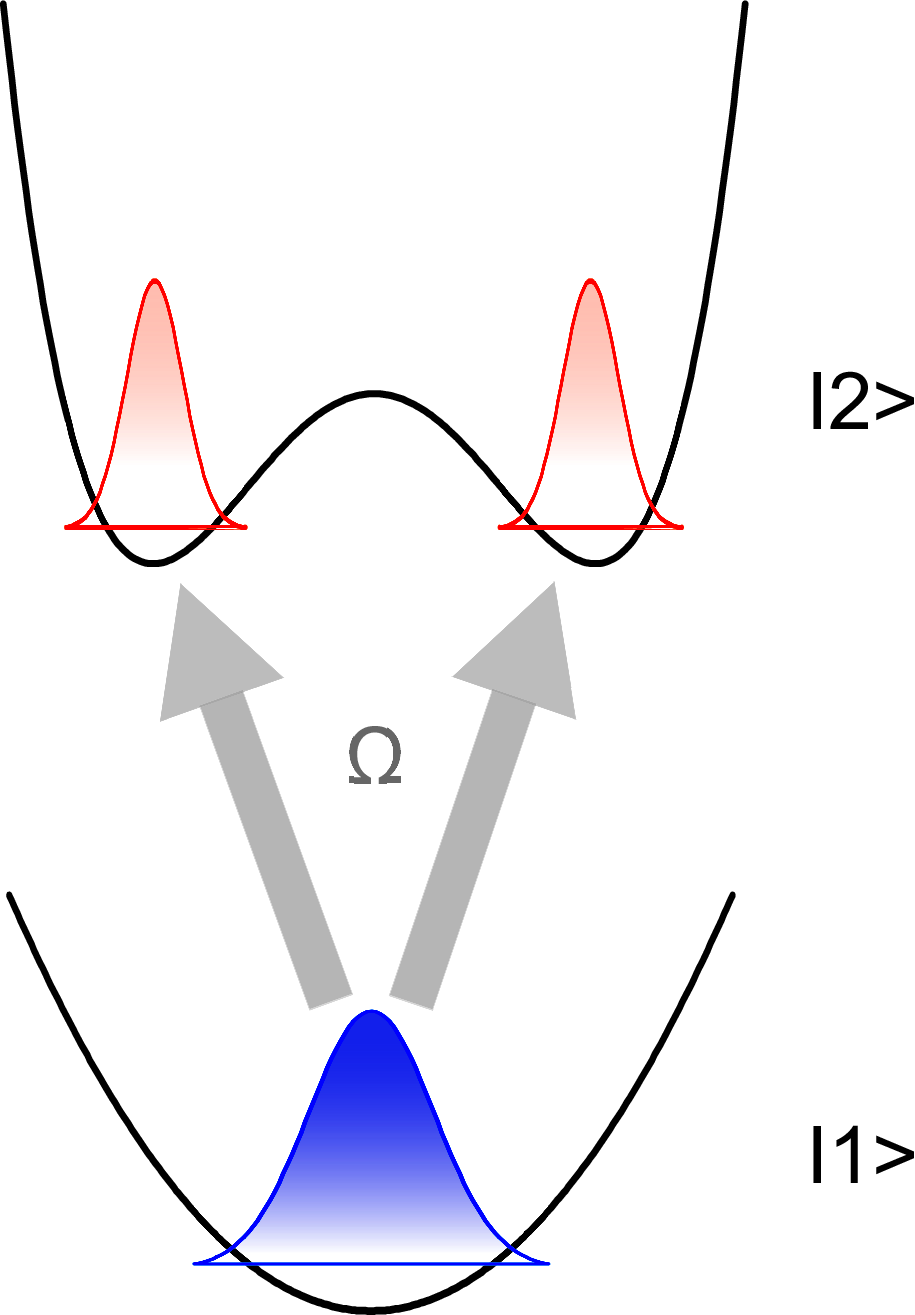}%
\caption{Schematics of the beam-splitter: atoms are transferred from a single to a double-well by coherently coupling (with Rabi frequency $\Omega$) two long-lived internal states in the presence of a state-selective potential.} 
\label{fig:schema}
\end{figure}
In order to avoid both to excite the cloud and to compromise its phase coherence, here we propose a novel, robust beam-splitter in which the double-well potential is not gradually raised up but the atoms are instead transferred from a single-well potential to a double-well by a $\pi$-like pulse within two different long-lived internal states $\vert 1 \rangle$ and $\vert 2 \rangle$, as depicted in Fig.\ref{fig:schema}. This procedure allows the atoms to reach their final split state $\vert 2 \rangle$ without passing through a deformation of the trapping potential itself, thus avoiding intrinsic excitations. 

\begin{figure}
\centering
\includegraphics[width=0.45\textwidth]{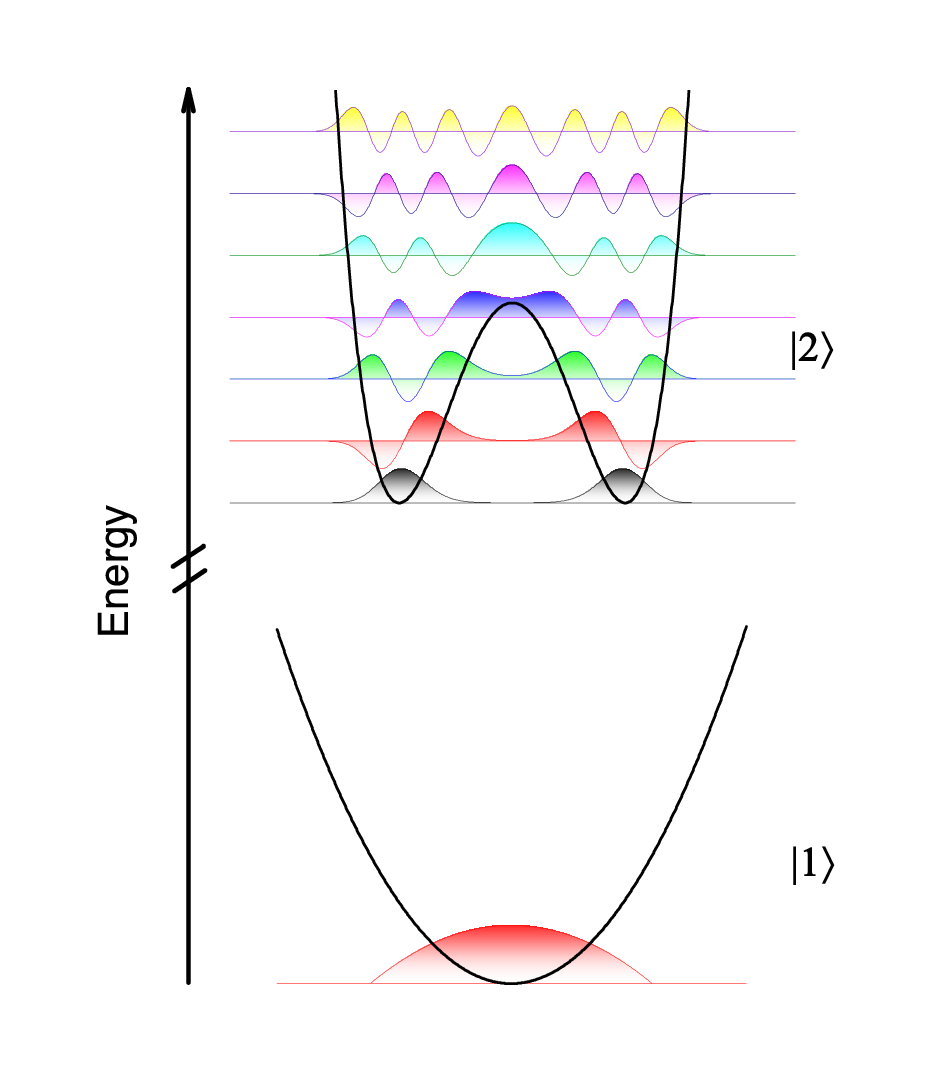}%
\caption{Real part of the symmetric wavefunctions for a BEC in the single trap and for different vibrational states in the double well. It appears clear that a better spatial overlap of the two external potentials in $\vert 1 \rangle$ and $\vert 2 \rangle$ would lead to a better matching of the initial BEC with the low lying energy states in the double-well. Higher energy separation between the eigenstates, $E_n$, and lower values of $\Omega$ reduce the contribution of the higher excited states in the transfer dynamics. The wavefunctions shown here refer to the set of parameters used in section II.A.} 
\label{fig:overlap}
\end{figure}

By using coherent electro-magnetic pulses to transfer the atoms to the double-well potential, we can address the exact final vibrational states which we want to populate, e.g. the ground state as required in this proposal, by properly tuning the frequency of the pulse.   
To perform a transfer to single eigenstates of the double-well potential, one additionally requires the duration of the pulse $\tau$ to be sufficiently long with respect to the timescales set by the confining potentials $\tau \sim \pi/\Omega \gg 1/\omega_{ho}$, where $\Omega$ is the frequency of the Rabi coupling and $\omega_{ho}$ is the smaller trapping frequency along the splitting direction. 

The phase coherence between the separated clouds in the internal state $\vert 2 \rangle$ does not depend on the details of the dynamics that they follow during the splitting but is ideally set by the Rabi coupling which phase-locks each of them to the atomic cloud in the other internal state $\vert 1 \rangle$. This guarantees the final split state to be prepared in a coherent superposition of two spatial modes which are localized at each well of a double-well potential in absence of significant direct inter-site tunnelling. This technique can be of great benefit for atom interferometry: the phase coherence between the spatial modes is preserved while the coherent coupling to the state $\vert 1 \rangle$ is active and is ideally decoupled from the dynamics of the splitting cloud. Moreover, realizing an ideal coherent state in a separated double-well potential would give the possibility to observe interaction induced squeezing, over-squeezing and the formation of phase cat states \cite{huang2006, piazza2006}. 

Depending on the ratio $\Lambda$ between the effective interaction energy and the Rabi frequency, different regimes are expected \cite{zibold2010}, in the approximation of identical spatial modes of the coupled states. In particular, for $\Lambda<1$ in the so-called Rabi regime the linear coupling is governing the time evolution and single-particle coherence is preserved \cite{boukobza2009}. For $\Lambda>1$ interactions start to play an important role, the Josephson regime is entered and for $\Lambda>2$ self-trapping modes (also with running-phase) appear. Atom transfer within internal states is progressively inhibited and also single-particle coherence is gradually lost \cite{boukobza2009}. An ideal coherent atom transfer would thus require $\Lambda \lesssim 2$. As $\Omega \ll \omega_{ho}/\pi$ to guarantee the absence of unwanted excitations during the process, the above condition is fulfilled only for small enough interaction energies. These are achievable in current experiments, for example, with the use of Feshbach resonances \cite{zibold2010}. Another possibility consists of dynamically varying the detuning of the Rabi pulse $\delta (t)$ in order to compensate for the effects of the interparticle interactions, as we will show.    

In the case we consider in this paper, the two spatial modes of the states $\vert 1 \rangle$ and $\vert 2 \rangle$ are different and their wavefunction overlap plays an important role in the transfer dynamics, see Fig.  \ref{fig:overlap}. This is responsible for the vibrational states that the atoms preferably occupy once they are transferred, for a given detuning of the pulse. It also strongly affects the effective time of the transfer and the overall fraction of atoms that can be moved to state $\vert 2 \rangle$, when dealing with interacting particles. 
In a simple single-atom picture, the amplitude of the transfer probability to the $n$th-vibrational state of the double-well potential $\vert 2, n \rangle$ can be written as $P_n=\frac{C_n\Omega}{\sqrt{(C_n\Omega)^2+\Delta_n^2}}$ where $C_n=\vert \langle 2, n \vert 1 \rangle \vert$ is the wavefunction overlap of the $n$th-state with the initial state $\vert 1 \rangle$ and $\Delta_n$ the pulse detuning to the $n$th-state, see also \cite{beaufils2011}. Thus, in order to mainly occupy the ground state ($n=0$), one requires $C_n \ll \Delta_n/\Omega$ for $n =1,2,3..$. This imposes a condition on the trap frequencies in $\vert 1 \rangle$, on the barrier height and separation of the minima of the double-well potential in $\vert 2 \rangle$, and on the Rabi frequency. In particular one configuration appears preferably suited for our proposal: the overlap with the ground state $\vert 1 \rangle$ is higher for the lower lying energy states in $\vert 2 \rangle$, i.e. $d/(2R)\lesssim 1$ with $d$ the minima separation in the double well and $R$ the Thomas-Fermi radius of the BEC in $\vert 1 \rangle$, and $\Delta_n/\Omega \gg 1$ to guarantee the occupation of the ground state with high fidelity. In the case the overlap is higher for the higher lying energy states in the double-well, $d/(2R)\gg 1$, and $\Delta_n/\Omega $ is large enough, occupation of the sole ground state can still take place but at the price of long transfer times, according to the low values of $C_0$. 

The absolute value of $\tau \sim \pi/\Omega$ should be indeed safely smaller than typical timescales of irreversible dephasing due to technical and intrinsic reasons, including instabilities of the magnetic fields and finite lifetime of the atomic samples in the traps. From previous experimental observations, we can estimate this limit for $^{87}$Rb atoms with typical trapping configurations to be given by the shortest lifetime in the usual clock states, which is $\sim 100$ ms.

\subsection{Numerical simulations}
\begin{figure}
\includegraphics[width=0.5\textwidth]{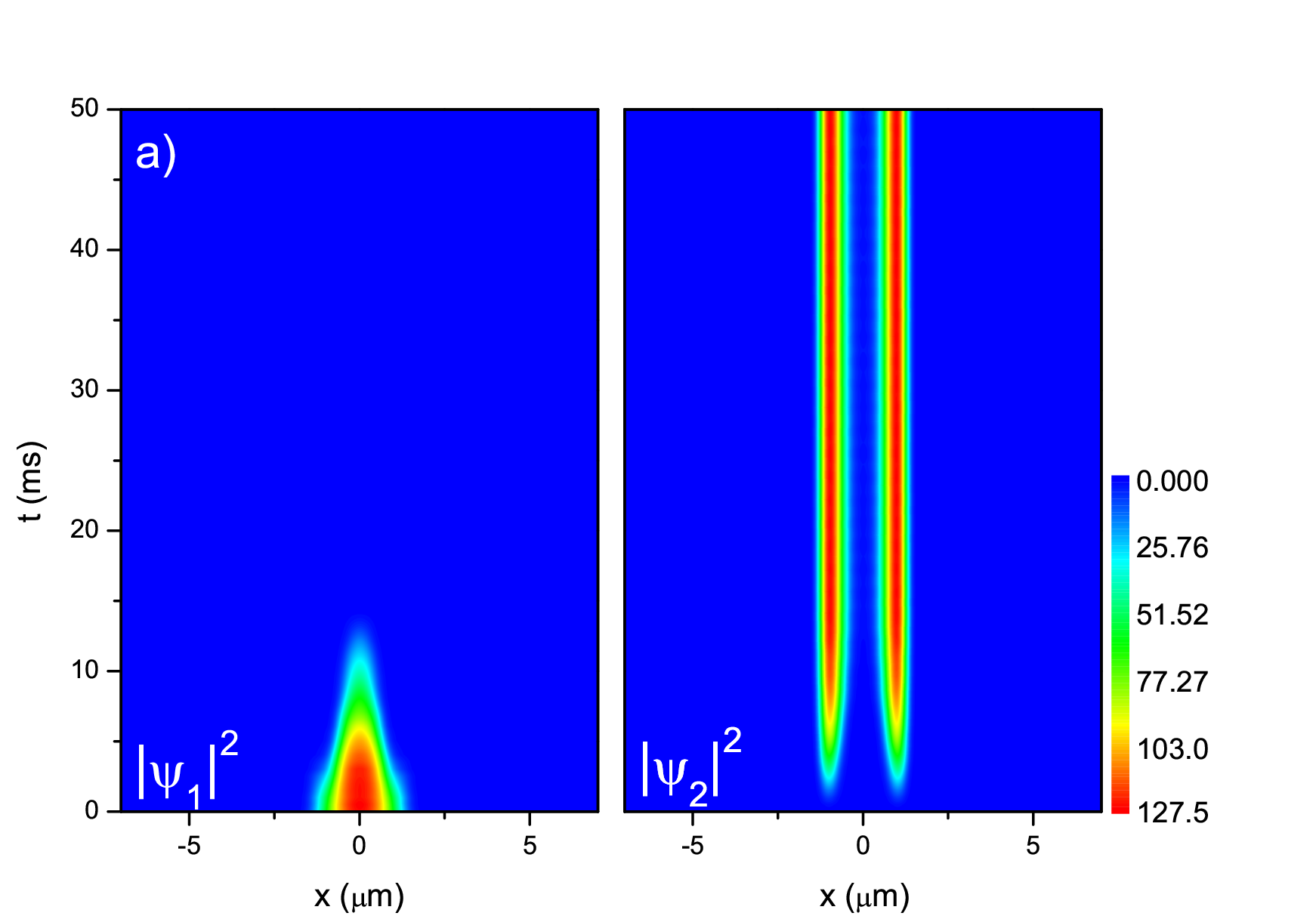}%
\includegraphics[width=0.45\textwidth]{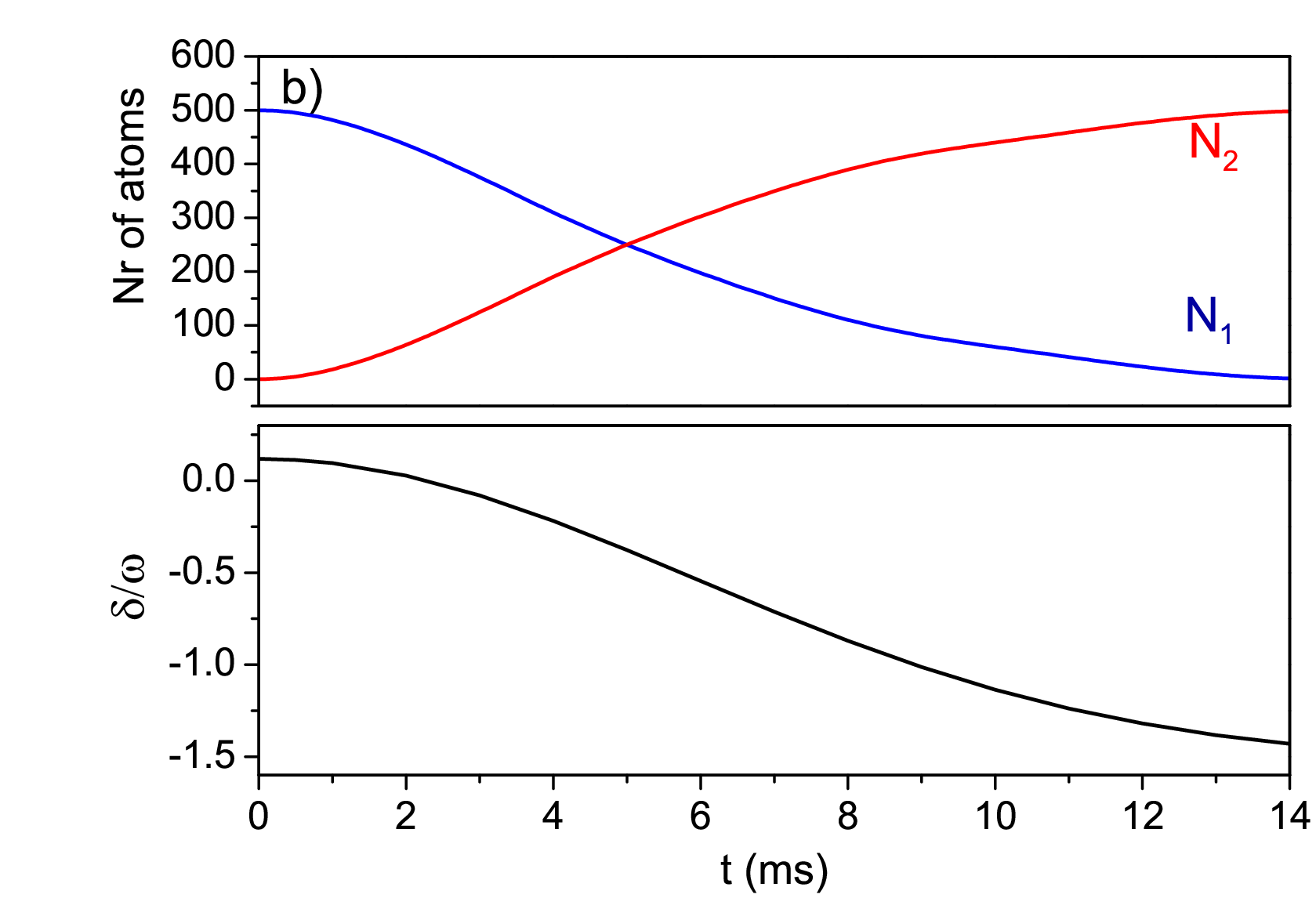}%
\caption{a) Density distribution of the atoms in $F=1$ (left) and $F=2$ (right) as a function of time. A Rabi coupling between the two states is applied for $14$ ms. b) In the upper panel the atom number $N_1$ and $N_2$ in the two states is shown as a function of time. The frequency of the pulse is numerically optimized to maximize the efficiency of the transfer, in the lower panel the detuning $\delta(t)$ is reported in units of the final trapping frequency.} 
\label{fig:theory}
\end{figure}

In order to more precisely characterize and to test our method, we numerically model the dynamics of two wave functions $\psi_1$ and $\psi_{2}$ with two coupled 1D Gross-Pitaevskii equations including the linear coupling due to the field driving Rabi oscillations:
\begin{eqnarray}
i \hbar \frac{\partial}{\partial t} \psi_1 &=&  \bigg[ - \frac{\hbar^2 \nabla^2}{2m} +V_1 + g_{11}\vert \psi_1 \vert^2 +g_{12} \vert \psi_{2} \vert^2 - \frac{ \delta}{2} \bigg]\psi_1  - \frac{\hbar \Omega}{2} \psi_{2}  \nonumber \\
i \hbar \frac{\partial}{\partial t} \psi_{2} &=&  \bigg[ - \frac{\hbar^2 \nabla^2}{2m} +V_2 + g_{22}\vert \psi_{2} \vert^2 +g_{21} \vert \psi_{1} \vert^2  + \frac{ \delta}{2} \bigg] \psi_{2} - \frac{\hbar \Omega}{2} \psi_{1} 
\label{eq:gpe}
\end{eqnarray}
where $m$ is the mass of the atom and $\delta$ the detuning with respect to the considered atomic transition \cite{zhang1994}. The effective one-dimensional interaction parameters are $g_{ij}=\frac{4\pi \hbar^2 a_{ij}}{m} \frac{1}{a_{\bot}^2}$ with $a_{ij}$ the intra-state ($i=j$) and inter-state ($i \neq j$) scattering lengths and $a_{\bot}$ the average harmonic oscillator length in the radial direction of the trap. Numerical calculations are performed by means of the time-splitting spectral method developed in \cite{wang2007}. In Fig. \ref{fig:theory} we show results for a BEC with $N=500$ atoms of $^{87}$Rb, where the internal states $\vert 1 \rangle $ and $\vert 2 \rangle $ correspond respectively to the hyperfine ground states $F=1$ and $F=2$. The initial trap $V_1(x)$ is harmonic with frequency $2\pi \times 500$ Hz and the double-well potential is approximated by the simple expression $V_2(x)=\frac{V}{(d/2)^4}(x^2-(d/2)^2)^2$ with barrier height $V/\mu_2=6.6$, where $\mu_2$ is the chemical potential in the double-well. The final trapping frequency is $\omega \simeq 2\pi \times 2100$ Hz. The distance between the two potential minima is taken to be $d=2$ $\mu$m. The frequencies in the radial direction are $2 \pi \times (50, 500)$ Hz, setting $d/(2R)=0.9$. We transfer the atoms from the $F=1$ to the $F=2$ ground state in $14$ ms with a pulse of Rabi frequency $\Omega=2\pi \times 100$ Hz ($\Lambda\sim1.9$). 
A compensation of the energy shift due to interactions is accomplished by varying $\delta$ to maximize the number of atoms transferred at each computational time step. In this way, by dynamically varying the detuning as shown in Fig. \ref{fig:theory}, we can transfer $100\%$ of the atoms in the internal state $F=2$. 
In order to characterize the fidelity of the wavefunction transfer \cite{trombettoni2008}, we calculate $f=\int dx \vert \psi_{2}^{*} (x,t)\vert \cdot \vert \psi_{2,gs}(x,t)\vert $, where $\psi_{2,gs}(x,t)$ in our case corresponds to the final ground state in $F=2$, calculated for the same atom number. For the case presented above the fidelity oscillates between $f=0.99994$ and $f=0.99998$, after the transfer has been completed. 
The system, which we have numerically studied, is an emblematic realization, with realistic experimental parameters, of the beam splitter which we propose in the present paper. Provided that the wavefunction overlap between the two coupled state is good, which we exemplify by the simple relation $d/(2R)\lesssim 1$, and that the ratio between the energy spacing of the low-lying eigenstates of the double well and the Rabi frequency is large enough, we have demonstrated that the ground state of the double well can be populated with excellent efficiency and fidelity. As a result no unwanted excitation have occurred during the transfer and the final state is ideally phase coherent.

\section{EXPERIMENTAL REALIZATION OF MW-DRESSED POTENTIALS}

\begin{figure}
\centering
\includegraphics[width=0.4\textwidth]{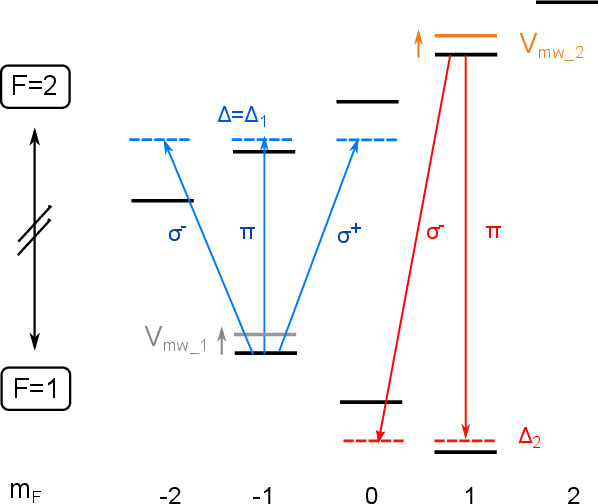}%
\caption{Level scheme of the $^{87}$Rb ground state in presence of an external magnetic field. The dressing transitions which we use in the present work are also indicated. The set of three arrows (blue) indicate the transitions which contribute to the dressing of $\vert 1, -1 \rangle$ with the polarization of the driving fields. For our experimental realization, in order to create a double-well in $F=1$ the MW-frequency is blue detuned by $\Delta=\Delta_1$ with respect to the $\vert 1, -1 \rangle \rightarrow \vert 2, -1, \rangle $ transition. Two arrows (red) indicate the dressing contributions for the $\vert 2, 1 \rangle$ state. In order to create a double-well the MW-frequency is instead red detuned by $\Delta_2$ with respect to the $\vert 2, 1 \rangle \rightarrow \vert 1, 1, \rangle $ transition.} 
\label{fig:system}
\end{figure}

\begin{figure}
\centering
\includegraphics[width=0.6\textwidth]{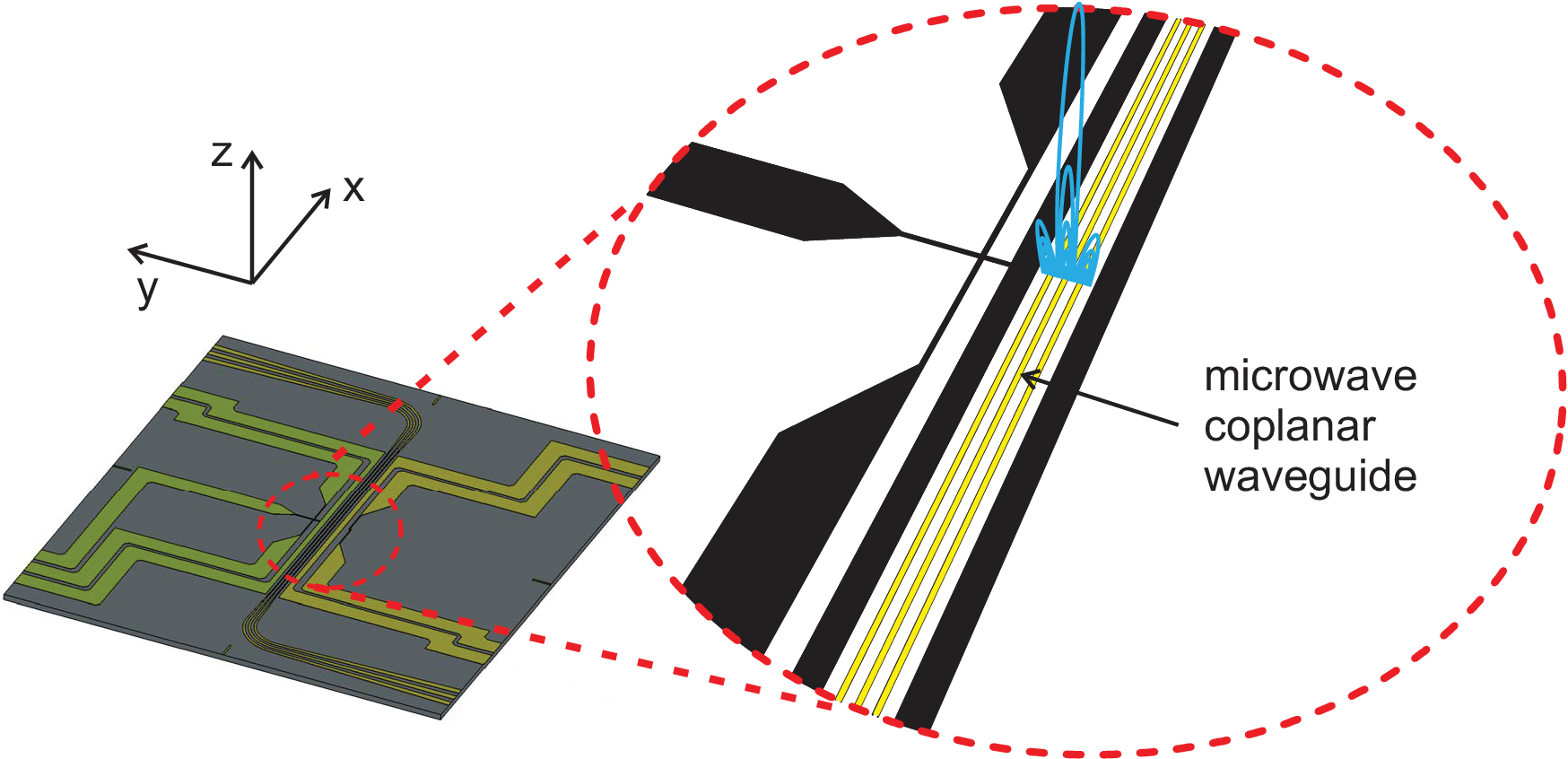}
\caption{Schematics of the chip first layer. A zoom in the centre of the chip shows the three wires which form the microwave coplanar waveguide. The central wire of the CPW is used, in addition, to carry the stationary current for the generation of the magnetic trap. In the present realization the two side wires of the CPW are grounded at the chip.} 
\label{fig:chip}
\end{figure}

The realization of our proposal requires the BEC sample to be transferred from a single-well potential to a double well, in a controlled way by a coherent pulse. Experimentally this is far from being a trivial task: the atoms should hence experience two different potentials (single and double-well), acting simultaneously on two different long-lived internal states which can be conveniently coupled by electro-magnetic radiation. The solution we adopt consists in coherently manipulating the BEC with state dependent microwave (MW) potentials. This technique has been employed before to selectively control the density distribution of $^{87}$Rb atoms in the clock states $\vert F=1, m_F=-1 \rangle$ and $\vert F=2, m_F=1 \rangle$ \cite{bohi2009,bohi2010}. The oscillating MW field $\textbf{B}_{mw}(\textbf{r})$ couples the different hyperfine states of the ground state via magnetic dipole transitions, see Fig. \ref{fig:system} for a scheme of the relevant levels and transitions. As a result of the action of the MW field which is continuously shone on the atoms, each state is dressed, i.e. is shifted in energy by the amount $V_{mw}(\textbf{r})$, effectively an AC-Zeeman shift which has a spatial dependence, see Fig. \ref{fig:system}. The MW induced potential can be approximated by $V_{mw}(\textbf{r})\approx  \sqrt{\vert \Omega_{1, m_1}^{2, m_2} \vert ^2+\vert \Delta_{1, m_1}^{2, m_2} \vert^2} - \vert\Delta_{1, m_1}^{2, m_2} \vert$ where: i) the Rabi frequency $\Omega_{1, m_1}^{2, m_2}(\textbf{r})$ depends on the projection of $\textbf{B}_{mw}(\textbf{r})$ along the direction parallel (normal), for linear (circular) polarizations, to the quantization axis set by the static magnetic field $\textbf{B}(\textbf{r})$ and ii) the detuning $\Delta^{2,m_{2}}_{1,m_{1}}(\textbf{r})=(\omega-\omega_{hfs})-(m_2-m_1) \mu_B \vert \textbf{B}(\textbf{r})\vert$, with $\omega_{hfs}$ the frequency of the hyperfine transition. This energy shift is varying in space through its dependence on the spatial mode of the MW and on the static magnetic field and it can be used to reshape the underlying magnetic potential used to trap the atoms. 
Moreover, in presence of a large enough Zeeman splitting of the hyperfine states, the MW-coupling can be effective only for some sublevels, leaving the others mainly unperturbed. This feature can be used to create state-selective, independently tunable potentials. The implementation which we have chosen has some advantages with respect to other state-selective techniques. Firstly, RF dressing techniques do not provide independent control on the different internal states and, secondly the use of optical potentials will suffer from spontaneous incoherent scattering by the trapping beams. 
As a first step towards the experimental realization of the beam splitter, we employ MW-dressing of the atomic levels, for the first time, to realize a double-well potential, selectively in one of the $^{87}$Rb internal states only. In order to do it we use an atom chip device with an integrated coplanar waveguide (CPW) for exploitation of the microwave near field. The design of the two-layer chip has been described in previous publications, see for example \cite{maineult2012}. We only recall here that our CPW is made of three gold wires parallel to the $x$-axis (see Fig. \ref{fig:chip} for axis orientation) each of size $(L_y, L_z)=(50, 10)$ $\mu$m and centres displayed by $150$ $\mu$m along the $y$ direction. The two side wires are grounded, whereas the central one is used to carry the dc-current which sets one of the confining axis of the Ioffe-Pritchard trap $\textbf{B}(\textbf{r})$. 
In order to use the spatial modulation of $V_{mw}(\textbf{r})$ to form a potential barrier in the centre of the trapping potential and thus realize a double well, three dominant effects can be exploited: the spatial dependence of the MW field $\vert \textbf{B}_{mw}(\textbf{r}) \vert$, of its polarization and the spatial dependence of the Zeeman shift due to the static field $\textbf{B}(\textbf{r})$. Thus the size of the achievable double-well potentials is strongly related to the geometry of the CPW and to its distance from the position of the atoms, as well as to the frequencies of the static magnetic trap. Notably, by playing with these parameters, splitting distances much smaller than the wavelength of the radiation employed can be achieved at a suitable distance from the chip. As an interesting additional feature, the great majority of the relevant fields in our experimental apparatus (MW for dressing, radio-frequency for two-photon transfer and part of the static magnetic fields for trapping) are delivered by the atom chip device within a compact setup.
 
\subsection{Characterization of the MW mode delivered by the chip}
\begin{figure}
\centering
\includegraphics[width=0.28\textwidth]{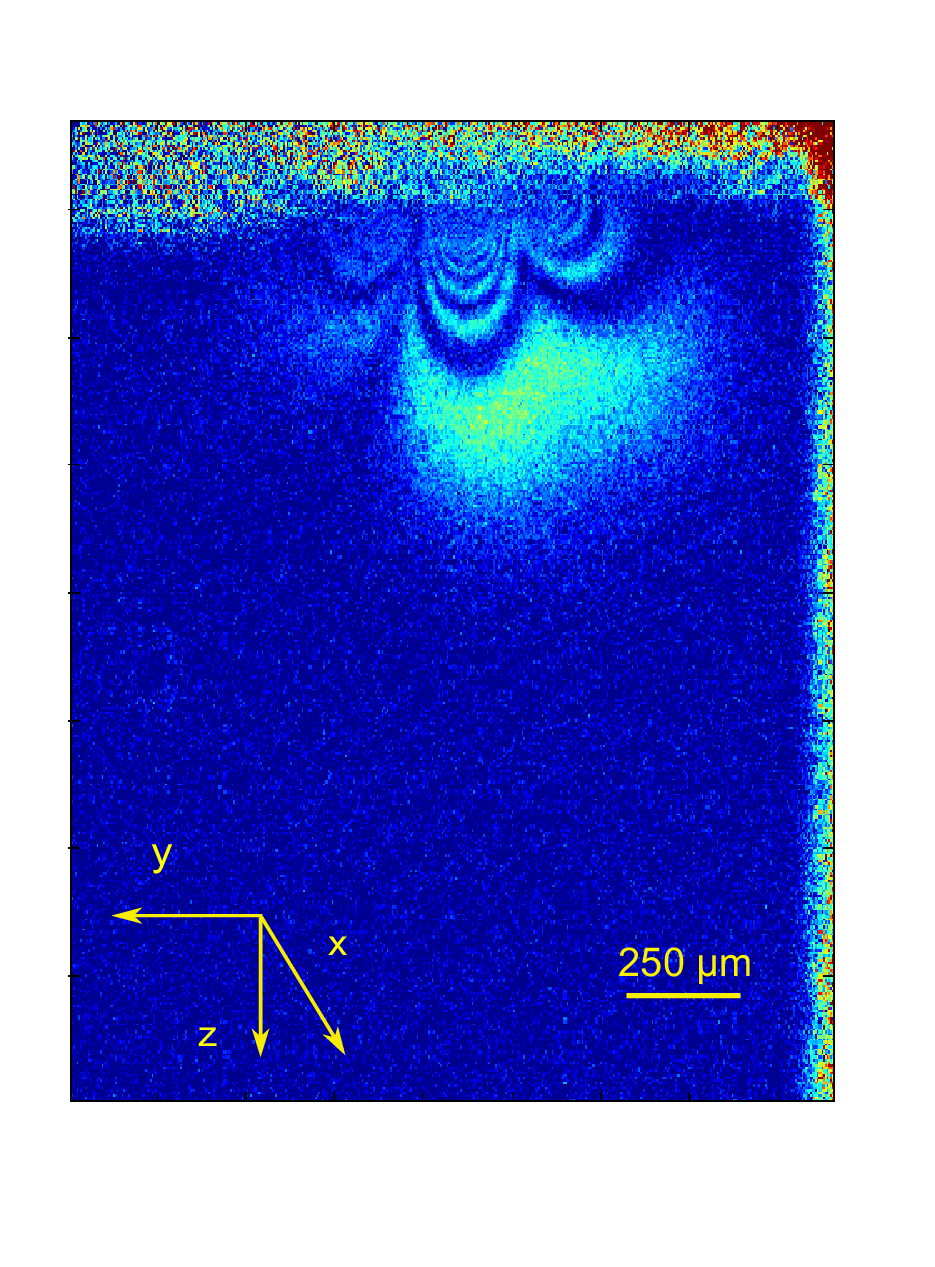}%
\includegraphics[width=0.5\textwidth]{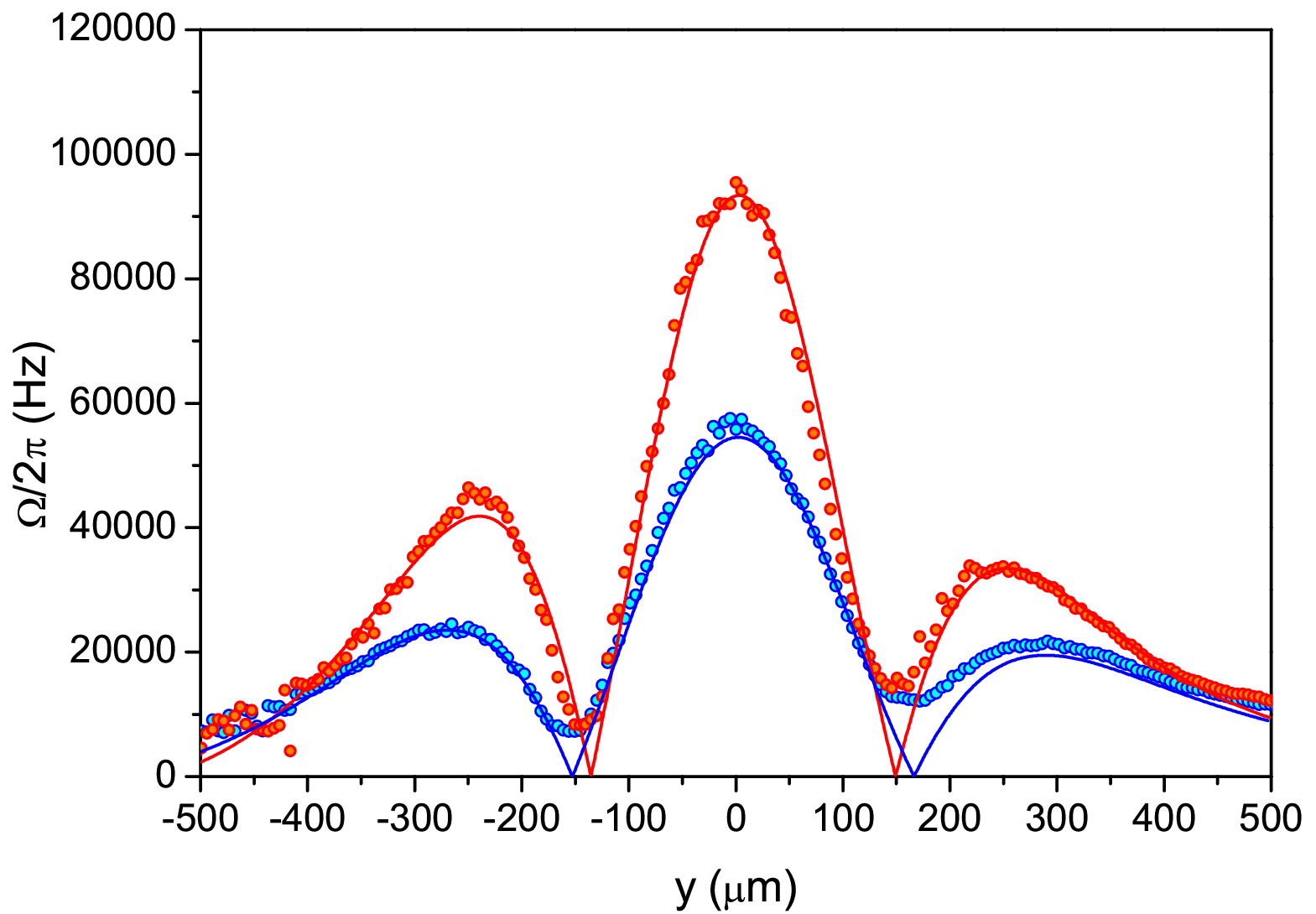}%
\caption{((Left) Atoms imaged in $F=2$ after a MW-pulse resonant on the $\vert 1, -1 \rangle \rightarrow \vert 2, -1 \rangle $ transition with power $10$ mW and $\textbf{B}\equiv \textbf{B}_{bias}//\textbf{y}$. The spatial modulation of the Rabi frequency is visible for a fixed duration of the pulse. On the upper side of the figure the shadow cast by the chip is visible. (Right) Points: measured Rabi frequency $\Omega_R(y,z_0)$ with MW-frequency resonant on the $\vert 1, -1 \rangle \rightarrow \vert 2, -1 \rangle $ and $\textbf{B}_{bias}//\textbf{y}$, here shown for two different distances from the chip $z_0=135$ $\mu$m (red points) and $z_0=185$ $\mu$m (blue points). Solid lines: corresponding simulations of the MW-mode.} 
\label{fig:system2}
\end{figure}

In order to identify the best strategy for the preparation of the double-well potential, we first characterize the microwave field delivered by the CPW similarly to what done in Ref. \cite{bohi2010}. We prepare a cloud of $2 \times 10^{6}$ atoms of $^{87}$Rb at a temperature of $T=10$ $\mu$K in the hyperfine sublevel $\vert F,m_F \rangle = \vert 1, -1 \rangle$. We then switch off all the laser beams and confining magnetic fields, letting the atoms expand in a static homogeneous magnetic field $B_{bias}\simeq3$ G, selectively pointing along the $x$, $y$ or $z$-axis. During expansion we pulse the MW-field with frequency resonant on the transitions $\vert 1, -1 \rangle \rightarrow \vert 2, m_F=-2,-1,0 \rangle$ and we selectively image the atoms in the $F=2$ hyperfine state, see Fig. \ref{fig:system2}(left). Varying the pulse duration, we can map the Rabi oscillation for the three different polarizations components with different orientation of the bias field. The intensity of the microwave field $B_{mw}(\textbf{r})$ and the relative weight of the different polarizations, in particular, are measured. We fit the MW-mode by means of a static field calculation. We can better reproduce the experimental data by considering an asymmetry of $0.16 I_{mw}$ in the current distribution of the external grounded wires of the CPW (ideally $I_{l}=I_{r}=-I_{mw}/2$ where $I_{l}, I_r, I_{mw}$ are respectively the currents in the left, right and central wires of the CPW, see also reference \cite{bohi2009}) and by assuming an induced current with amplitude $0.1 I_{mw}$ in one of the closest wires to the CPW. One example of the measured Rabi frequency, on the transition $\vert 1, -1 \rangle \rightarrow \vert 2, -1 \rangle$ ($\pi$ polarization) with $\textbf{B}_{bias}//\textbf{y}$, is shown in Fig. \ref{fig:system2}(right) together with the corresponding simulation, for two different distances from the chip surface.     

\subsection{Realization and characterization of a double-well potential on an atom chip}
\begin{figure}
\centering
\includegraphics[width=0.8\textwidth]{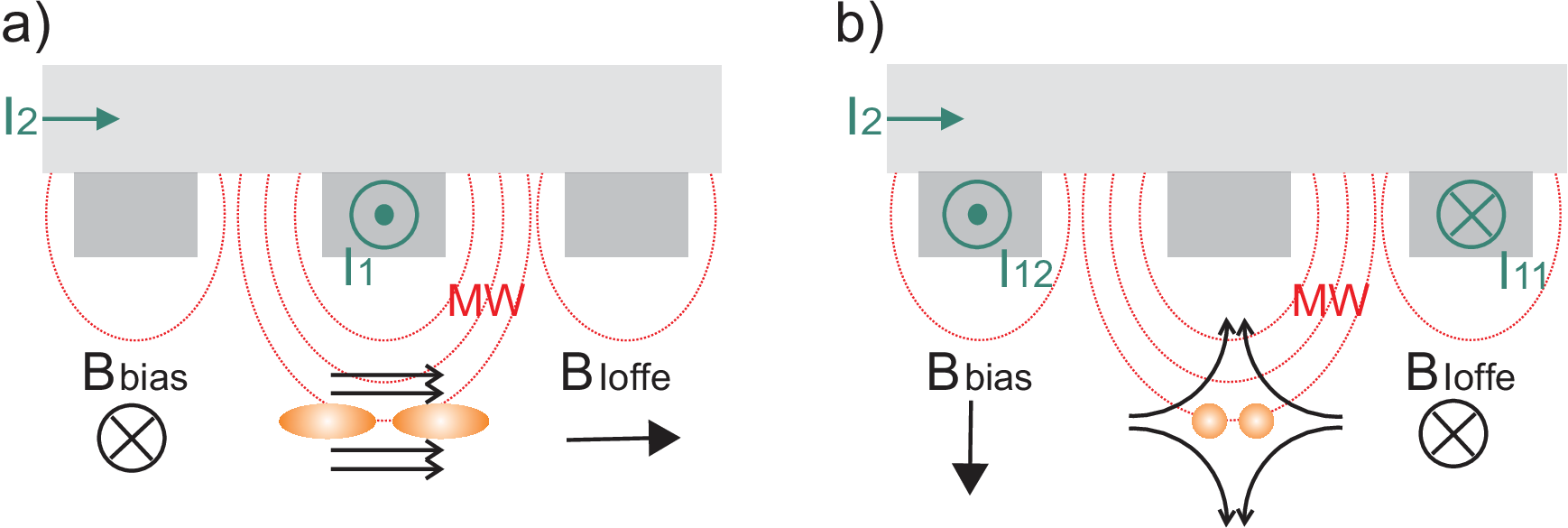}%
\caption{Schematics of the two-layer chip with the main fields for the realization of the MW dressed double well potentials. The currents $I_1$ ($I_{11}, I_{12}$) and $I_2$ flow in dedicated wires in the first and second layer of the chip respectively and contribute together with the Ioffe magnetic field $\textbf{B}_{Ioffe}$ and a bias field $\textbf{B}_{bias}$ to the formation of the static magnetic trap where the atoms are kept. The MW field irradiates from the CPW, which is made of three parallel wires as sketched in the figure, see also Fig. \ref{fig:chip}. a) Due to the action of the MW a double well is created along the longitudinal axis of the trap for the two states $\vert1,-1 \rangle$ and $\vert 2,1 \rangle$ when the MW frequency is tuned close to the $\pi$-polarization driven transitions as shown in Fig. \ref{fig:system}. b) A double well is formed along the radial axis of the trap, for MW frequencies close to $\sigma$-polarization driven transitions. In the two cases the direction of the splitting is set by the orientation of the MW field at the position of the atoms. } 
\label{fig:wires}
\end{figure}

As outlined above, both the MW-field and the static magnetic field of the trap can be used to locally change the dressing $V_{mw}$. Given the rather smooth spatial variation of our MW-field on the typical sizes of the atomic samples, we can additionally exploit the effect due to the static magnetic trapping to obtain double-well configurations with reduced minima separations. In this case, depending on the detuning of the MW field with respect to the atomic transitions the dominant contribution to $V_{mw}$ can thus be due to the spatial dependent Zeeman shift $\vert \textbf{B}(\textbf{r}) \vert$ or to the gradient of the MW field polarization with quantization axis set by the magnetic field of the trap $\textbf{B}(\textbf{r})/\vert \textbf{B}(\textbf{r}) \vert$. With our chip design, the MW-field delivered by the CPW can be used to create a double-well potential for the atoms in each of the internal states $\vert F=1, m_F=-1 \rangle$ or $\vert F=2, m_F=1 \rangle$ with two different possible realizations of the static trap, see Fig. \ref{fig:wires}. In the first case the Ioffe trap is oriented with its weak axis orthogonal to the axis of the CPW, i.e. $\textbf{B}_{Ioffe} //\textbf{y}$ see Fig. \ref{fig:wires}a. The MW-field will lead to the formation of a double-well along the axial direction of the trap, when its frequency is tuned close to the $\pi$-induced transitions ($\Delta m_F=0$), see the transitions highlighted in the scheme of Fig \ref{fig:system}. The sign of the detuning $\Delta$ is set in order to produce a repulsive potential $V_{mw}>0$ in the centre of the magnetic trap and it is thus blue(red)-detuned for the dressing of the $F=1(2)$ states.
In the second case the weak axis of the trap is parallel to the axis of the CPW, i.e. $\textbf{B}_{Ioffe} // \textbf{x}$ as shown in Fig. \ref{fig:wires}b. In this case the double-well is formed along the radial direction of the trap, $\textbf{y}$ direction in the figure, by blue-detuning the MW-field close to the $\sigma^{-}$-induced transition ($\Delta m_F=-1$ and $\Delta<\mu_0B_{Ioffe}/\hbar$) for the atoms in $F=1$. The atoms in $F=2$ can be instead dressed by setting the MW-frequency blue-detuned with respect to both the $\sigma^{-}$ and the $\pi$-induced transitions. This latter configuration can be used to obtain an experimental implementation of the simulations presented in the previous paragraph. However, in the current version of the experiment we are limited to the unique realization of the first case described here. Indeed, the only wires that we could use to generate the magnetic trap shown in Fig. \ref{fig:wires}b, are the side wires of the CPW which we cannot independently control as they are grounded together at the chip. Conversely, a new version of our experiment, which is now under construction, will meet these requirements. We finally notice that the splitting direction is set by the orientation of the MW field at the position of the atoms, thus, in order to modify it, a different design of the CPW would be required.   

\begin{figure}
\centering
\includegraphics[width=0.4\textwidth]{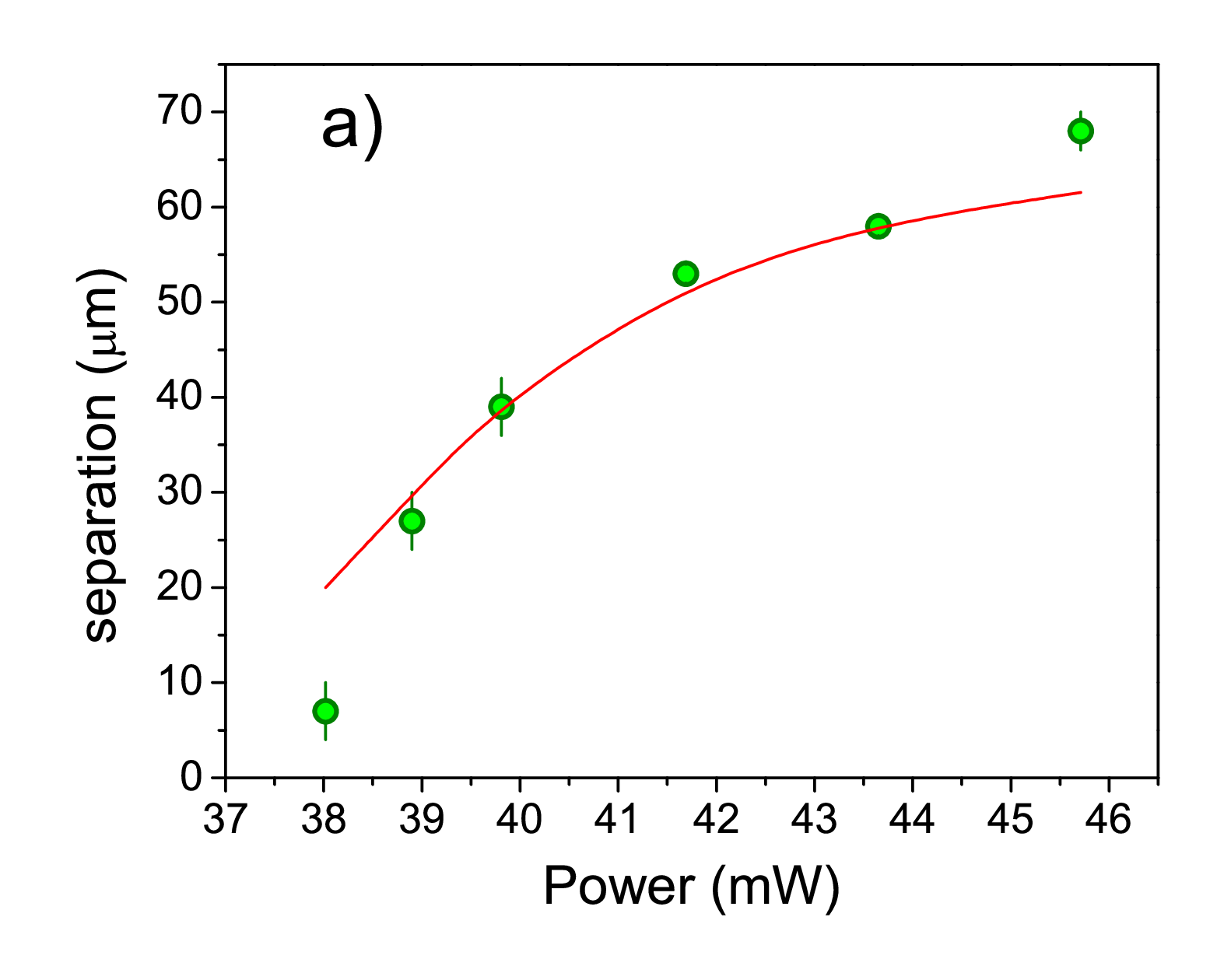}%
\qquad\qquad
\includegraphics[width=0.4\textwidth]{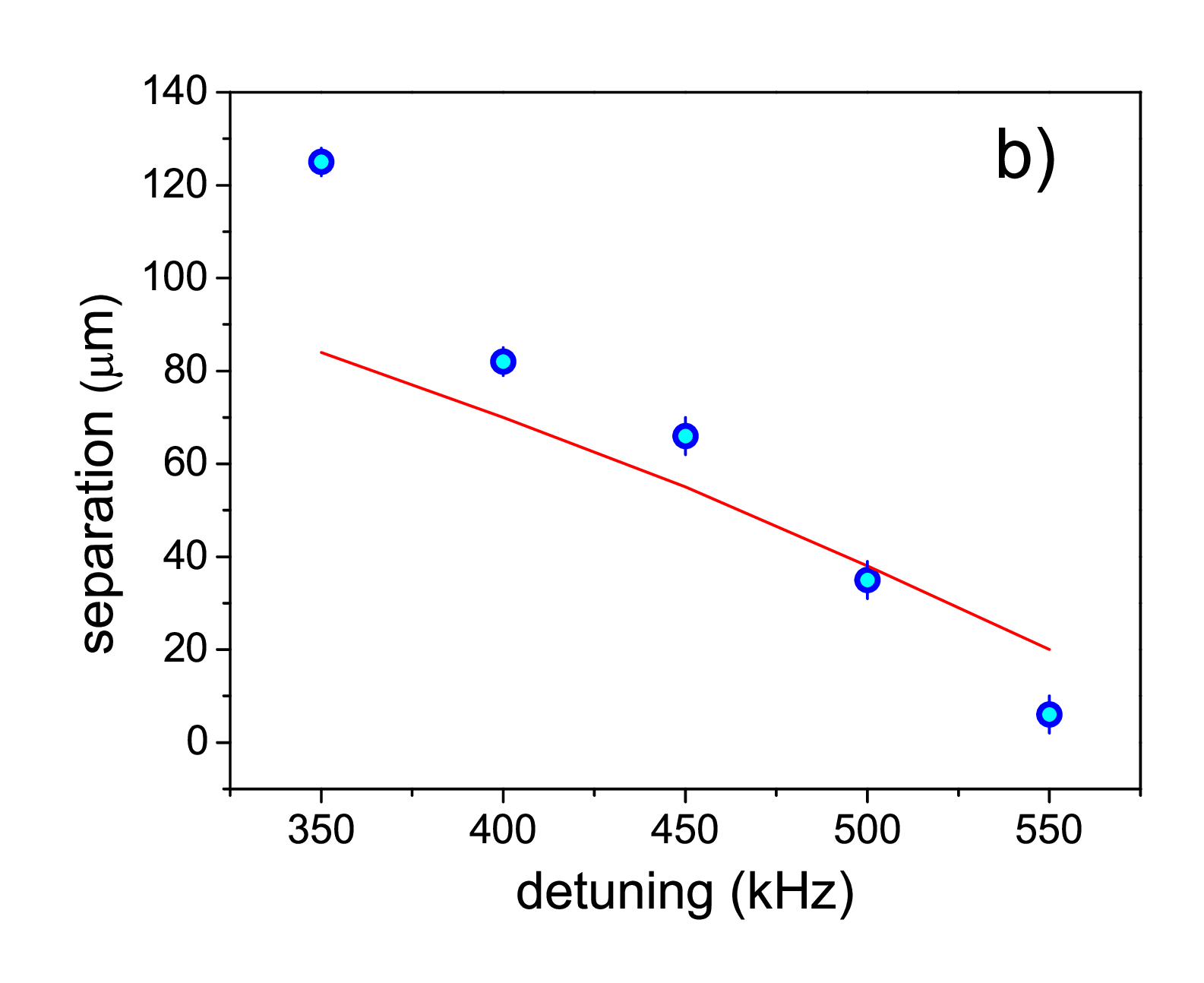}
\includegraphics[width=0.4\textwidth]{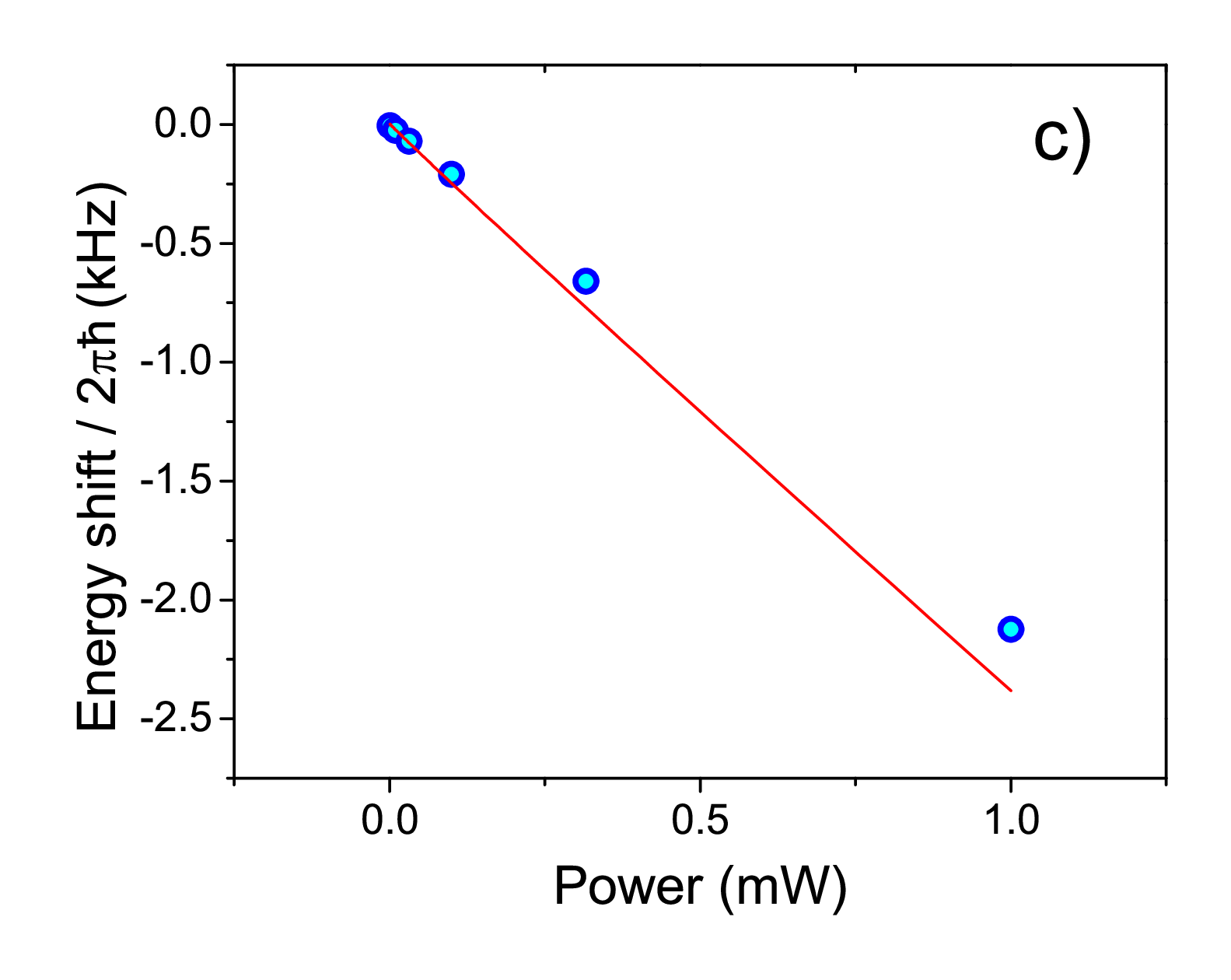}%
\caption{ Dressing of $F=1$ state. a) Measured cloud separation as a function of the power delivered by the synthesizer with $\Delta=500$ kHz. Solid line represents the simulations without free parameters. b) Measured cloud separation as a function of the detuning $\Delta$ for $P_{mw}=39$ mW with corresponding simulations. c) Energy shift of the transition $\vert F=1, m_F=-1 \rangle \rightarrow \vert F=2, m_F=1$ as a function of the power delivered by the synthesizer for $\Delta=500$ kHz together with simulations.} 
\label{fig:dressing}
\end{figure}

In order to test the predictions and characterize our system, we first show experimental results obtained by dressing just one of the two ground states, $\vert 1, -1 \rangle$. By using the microwave field generated by the CPW, we can create a double-well potential with MW frequency detuned to the blue, by a suitable amount $\Delta$, with respect to the transition $\vert 1, -1 \rangle \rightarrow \vert 2, -1\rangle$, as explained above (blue arrow in the scheme of Fig. \ref{fig:system}). The resulting dressed potential is approximately a sum of the three different contributions from all the possible transitions $V_{mw}=V_{2,-2}+V_{2,-1}+V_{2,0}$, where only the $V_{2,-1}$ term is leading to the formation of a double-well, whereas the others mainly shift the vertical position of the cloud. The final potential to the atoms is characterized experimentally by measuring the distance between the potential minima as a function of the microwave amplitude delivered by the synthesizer. We prepare a Bose-Einstein condensate of $\sim 1000$ atoms in the state $\vert 1, -1 \rangle$. For the experiments described here, the atoms are loaded in a harmonic trap with frequencies $2 \pi \times (160, 90, 130)$ Hz and longitudinal axis along the $\textbf{y}$ direction, at $50$ $\mu$m distance from the chip surface. We switch on adiabatically the microwave with a s-shaped ramp in $ \sim 1$ s and we measure the density distribution of the atoms after a short time-of-flight (TOF) of few ms. For these measurements, the microwave frequency is set to be $\Delta=500$ kHz blue-detuned with respect to the $\vert1,-1 \rangle \rightarrow \vert 2,-1 \rangle$ transition, with a Ioffe field of $B_{y}=3.055$ G. Increasing the power of the MW, two separated clouds start to be resolved with increasing separation distance. To correct for an asymmetrical atomic population of the two well, mainly due to the underlying asymmetry of the microwave field, a constant magnetic gradient of $68$ mG/cm is added. 
In Fig.\ref{fig:dressing}a we compare our results to numerical simulations performed by taking into account the actual geometry of the chip wires with no free parameters. The agreement with the measurements is quite good. In Fig.\ref{fig:dressing}b the measurement of the separation distance is shown as a function of the detuning $\Delta$, for a power of $38.9$ mW. A non-negligible effect of the MW near-field consists also in pushing the atoms away from the chip due to the presence of a gradient of $\vert \Omega_{2,-1}(\textbf{r}) \vert$ along the $z$-axis. Finally we measure the shift of the resonance of the $\vert1,-1 \rangle \rightarrow \vert 2,1 \rangle$ transition for increasing power of the MW field, which we compare to the calculated energy shift induced by the dressing on the coupled levels, Fig.\ref{fig:dressing}c. The transition is driven by electro-magnetic pulses of a MW field, $500$ kHz blue detuned from the $\vert 2, 0 \rangle$ level, and an additional RF field with power low enough not to introduce additional dressing. The agreement with theory is generally good for relatively small splitting distances $d \lesssim 50$  $\mu$m. The residual disagreement is to be attributed to the simple modelling of the MW-mode (we estimate local deviations up to $20\%$, whose effect becomes more important for higher MW powers) and, to a lower extent, of the actual chip trap. This allows us to extrapolate from the simulations, in this range, the functional dependence of double-well barrier height on the splitting distance, which is only smoothly affected by the precise detuning $\Delta_{1,-1}$ chosen, being predominantly set by the magnetic trap frequencies and by the MW-field mode. For our experimental case, we find that to completely split a condensate with chemical potential $\mu/(2 \pi \hbar) \sim 100$ Hz, a separation $\gtrsim 30$ $\mu$m should be achieved. This rather large separation is due to the smooth variation of $\textbf{B}_{mw}$ over the size of the cloud and to the low trapping frequency along the splitting axis, which coincides with the longitudinal axis of the trap. To obtain a separated double well with smaller splitting distances of a few microns, the CPW transverse dimension could be reduced by an order of magnitude (as in Ref. \cite{bohi2009}) or the trapping frequency could be increased, as in the second case discussed above where the splitting is taking place along the radial axis of the trap. 
   
\section{REALIZATION OF STATE-SELECTIVE POTENTIALS ON A CHIP}
In the previous section, we have shown that we can use a chip-delivered MW-near field to create a double-well potential, alternatively to the so far employed radio-frequency fields \cite{hofferberth2006}.
Moreover a key feature of MW-generated dressed potentials with respect to radio-frequency ones relies on the possibility of independently addressing the internal states used in the beam splitter proposal. The RF field will indeed couple to both the states, limiting the possibilities of independently engineering the wavefunctions of the atoms in the two states. As discussed above, the overlap between the wavefunctions plays an important role in determining the feasibility and fidelity of the beam splitter operation and it is thus a crucial parameter to be optimized. The MW field, notably, allows to create state-selective potentials, which can also be independently tuned once two different frequencies are provided. 

In order to test this, we dress the atoms by superimposing on the CPW two MW-frequencies delivered by the same synthesizer, which can be independently tuned. In particular we add to the dressing on the $\vert 1,-1 \rangle \rightarrow \vert 2,-1 \rangle$ transition, which affects mainly the atoms in $F=1$, a second frequency close to the $\vert 1,1 \rangle \rightarrow \vert 2,1 \rangle$ transition for the dressing of the atoms in $F=2$, as shown in the level scheme of Fig.\ref{fig:system}. Note that due to the Zeeman splitting following the presence of a bias field of $\sim 3.055$ G, the two frequencies are far enough ($\sim 9.2$ MHz) not to give rise to cross interference terms. 

\begin{figure}
\centering
\includegraphics[width=0.55\textwidth]{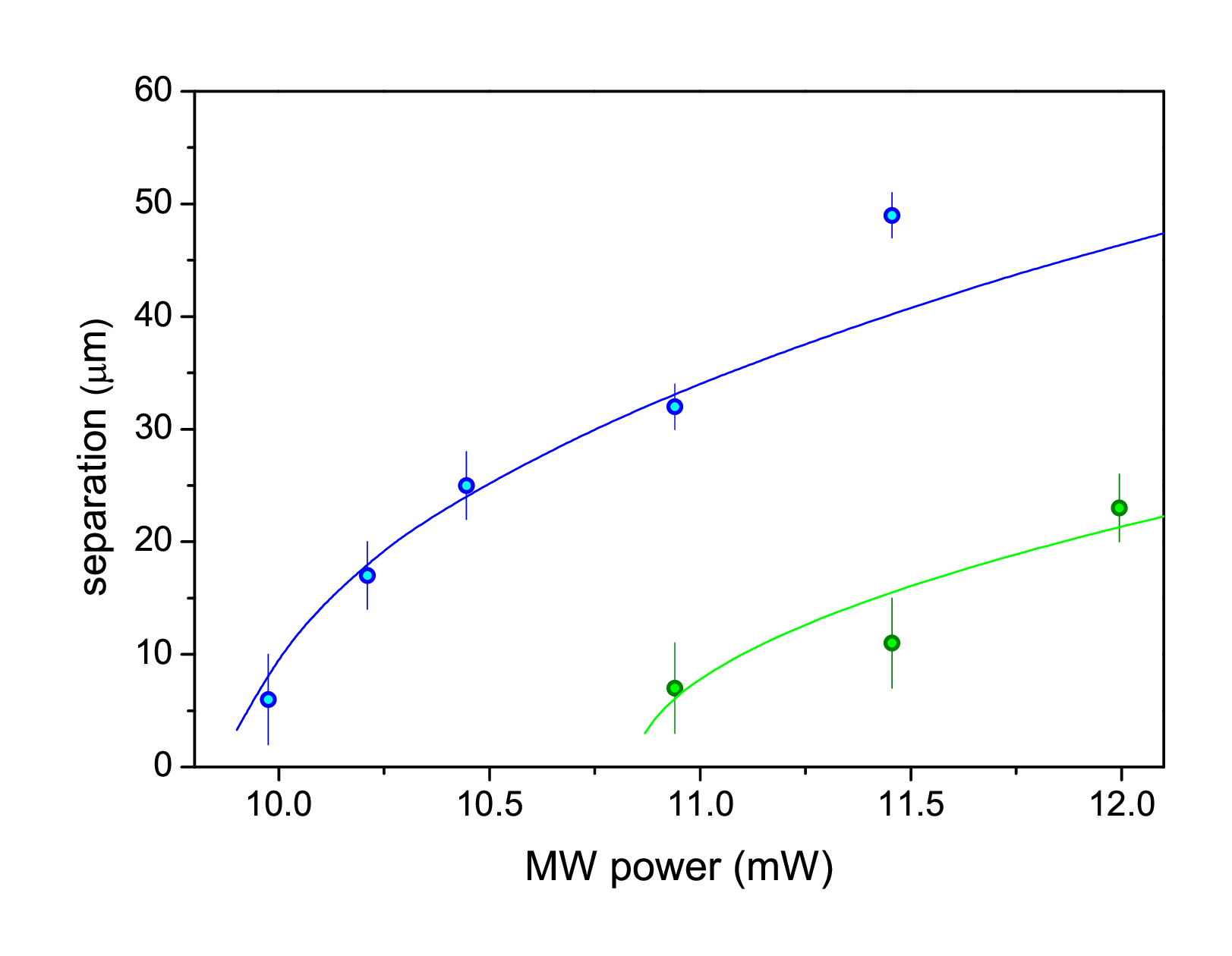}%
\qquad\qquad
\includegraphics[width=0.33\textwidth]{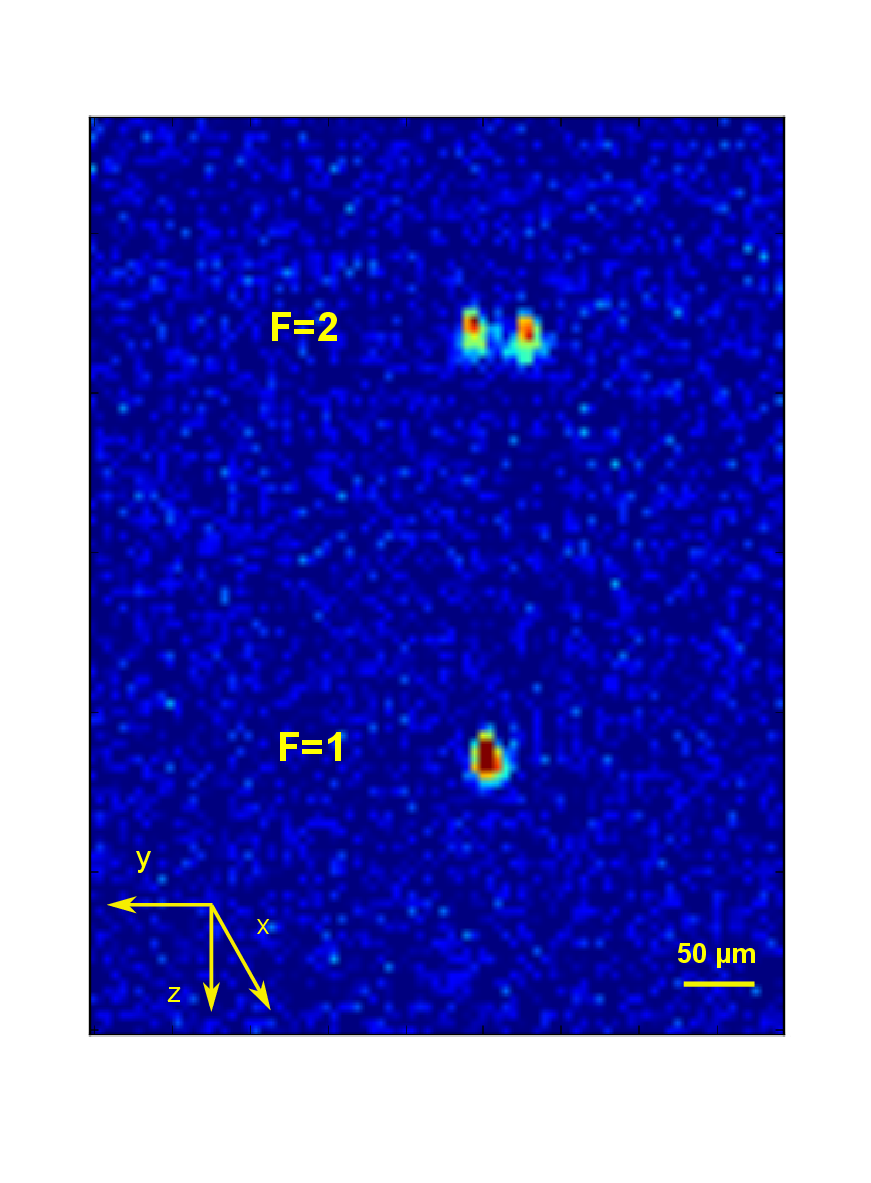}
\caption{ Left: simultaneous two-tone dressing of $F=1$ and $F=2$ with $\Delta_{1}=-\Delta_{2}=52$ kHz and equal power delivered by the synthesizer in the two tones. The atoms are all either in $F=1$ or in $F=2$, and they are detected at the same TOF. The solid lines are fits to the data obtained by using the numerical simulations of the fields with the mw amplitude as a free parameter. Right: double state detection of the atoms in $F=1$ and $F=2$. Here the atoms are prepared in the two states and simultaneously dressed by a two-frequency MW radiation. Vertical separation is due to the different time at which the imaging is performed for the two states ($15$ ms and $17$ ms TOF respectively).} 
\label{fig:twotonedress}
\end{figure}

We prepare a BEC alternatively in the state $\vert 1,-1 \rangle$ or $\vert 2,-1 \rangle$ and we ramp up the two-tone dressing adiabatically. In Fig.\ref{fig:twotonedress} we report the measurements at short TOF of the splitting distance when a double well is formed, with detuning $\Delta_1=-\Delta_2=52$ kHz and power $P_{mw1}=P_{mw2}$. A clear double well is forming in the $F=2$ state at lower powers than in $F=1$. This difference is due to the presence, in the dressed potential of the $F=1$ state, of an additional repulsive component on the $\sigma^{-}$-induced transition $V_{2,-2}$ which shifts the atoms along the $z$-axis to lower values of the MW potential $V_{mw,1}(\textbf{r})$. The magnetic gradient which is used to symmetrize the atomic distribution within the two wells is optimized to the slightly lower value of $57$ mG/cm for the atoms in $F=2$. As shown in Fig.\ref{fig:twotonedress}(left) a working configuration can be easily found in which the atoms in $F=2$ are well split into a double-well while the atoms in $F=1$ are still trapped within a single well. Dressing of $F=1$ state can be further independently manipulated in order to adjust the shape and position of the initial cloud.

\subsection{Transfer and spectra} 
\begin{figure}
\centering
\includegraphics[width=0.8\textwidth]{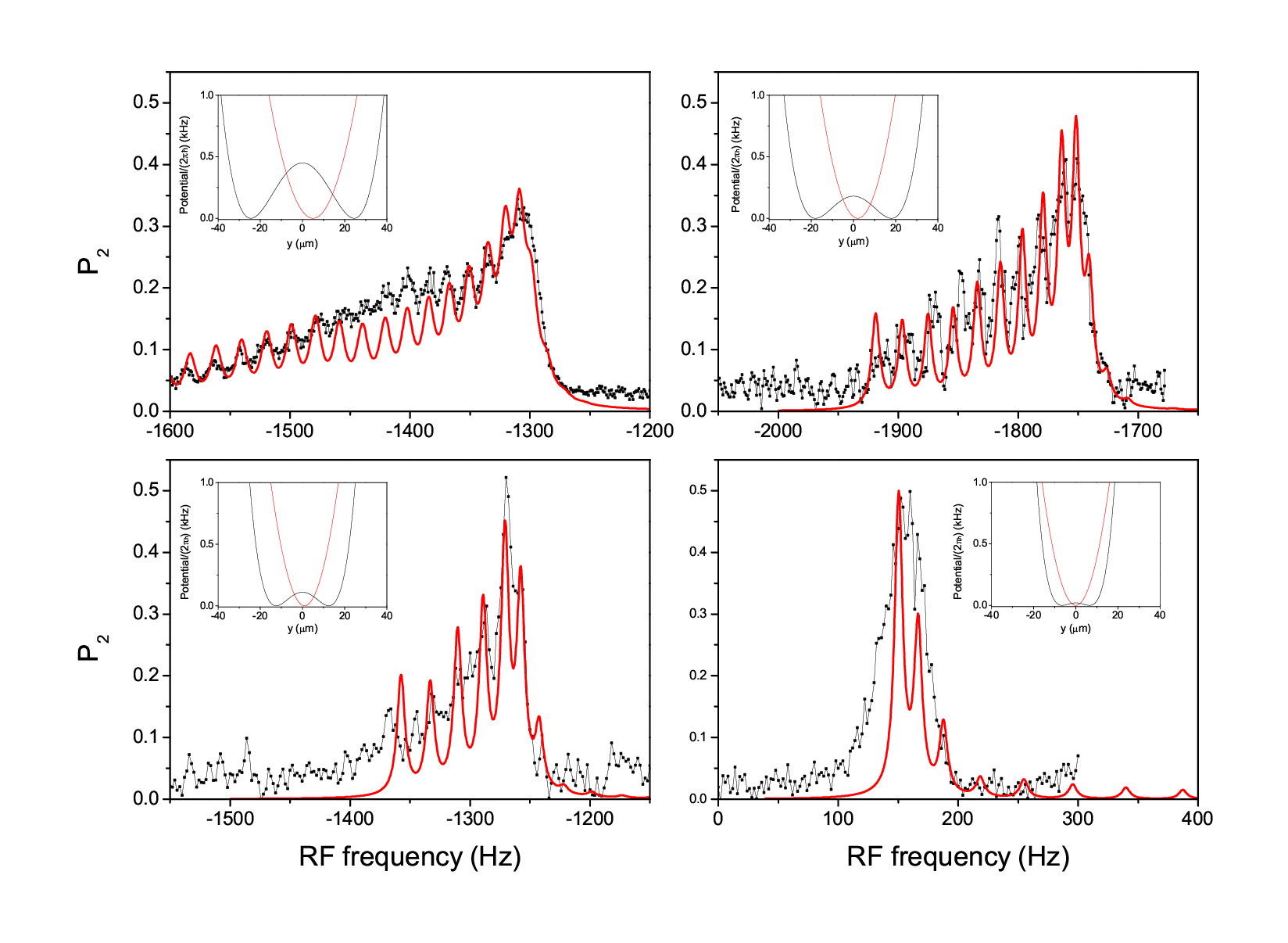}%
\caption{Relative number of transferred atoms $P_2$ from $\vert 1, -1 \rangle$ to $\vert 2, 1 \rangle $ for different powers of the dressing field $P_{mw}=11.5, 11, 10.5, 10$ mW, corresponding to different depths (and minima separations) of the double-well potential. These measurements are the average over $(20, 10, 5, 10)$ acquisitions respectively. Red solid lines are fit with a single-particle 1D theory (see main text for details). Insets show the potentials in $F=1$ (red solid line) and $F=2$ (black solid line) which correspond to the fits (see text). } 
\label{fig:spectra}
\end{figure}

\begin{figure}
\centering
\includegraphics[width=0.45\textwidth]{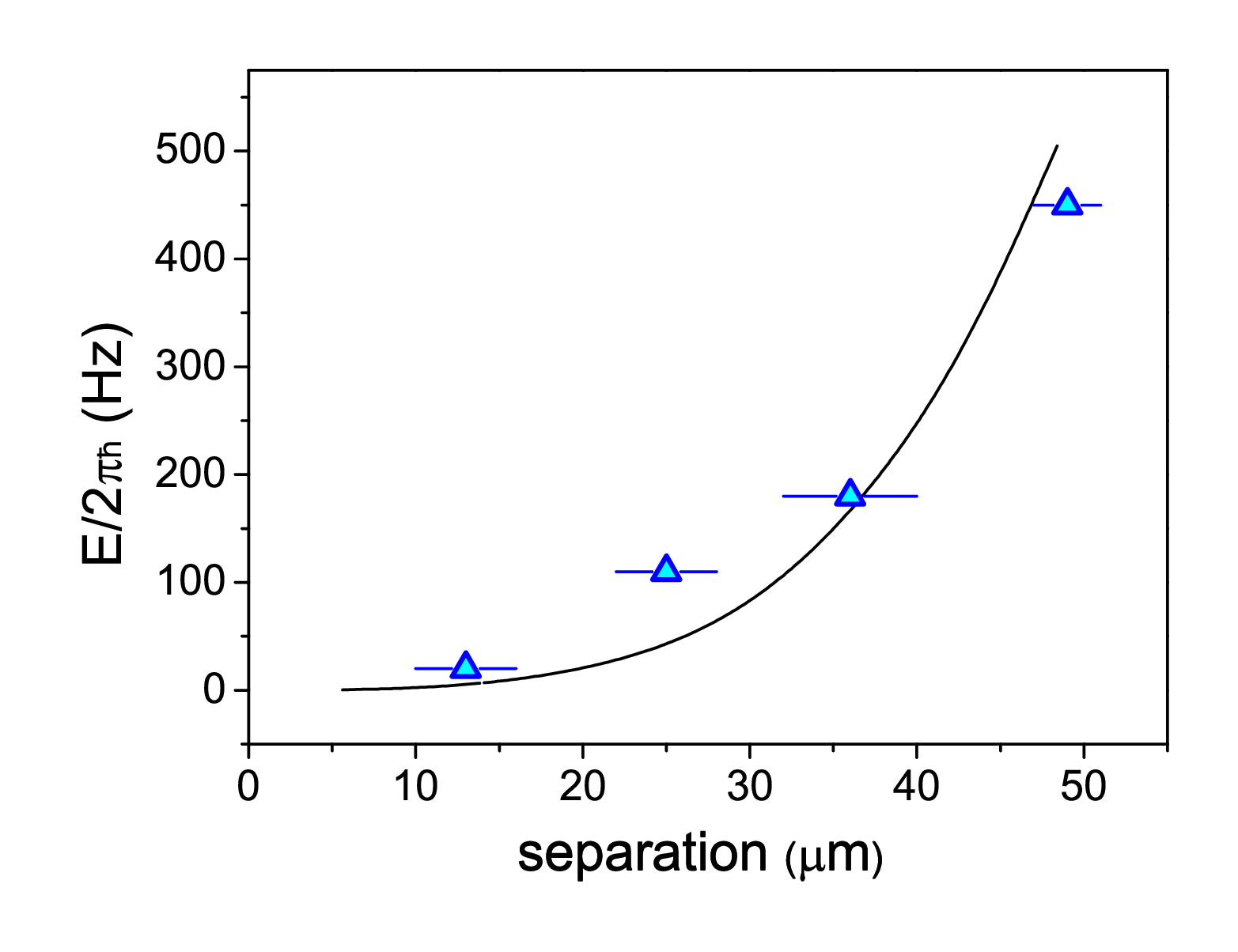}
\qquad\qquad
\includegraphics[width=0.45\textwidth]{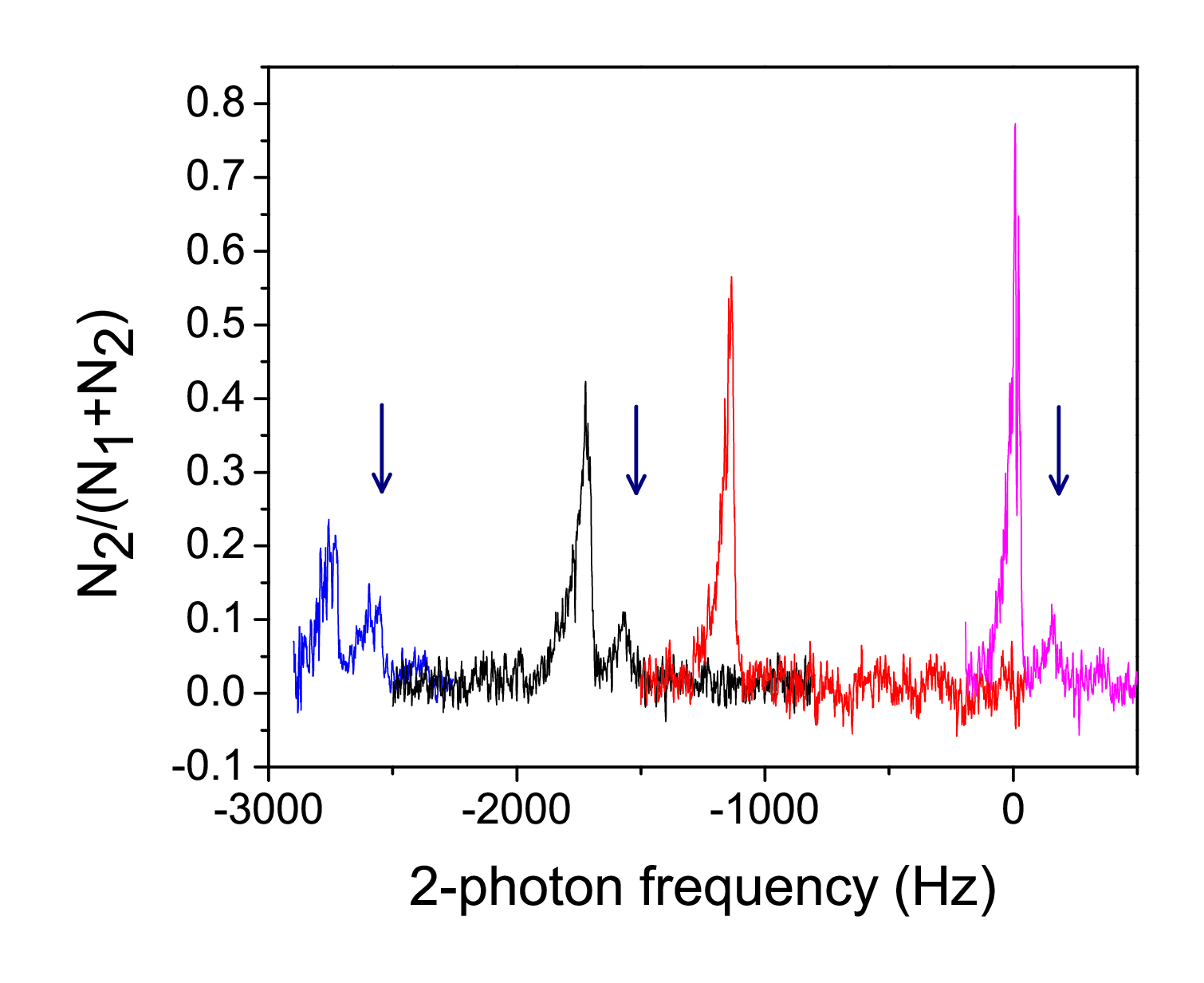}%
\caption{Left: Calculated energy height of the double-well barrier $E$ as a function of the separation between the minima of the potential. For comparison we plot the measurements of $E$, blue points, obtained by means of a simple fit of the spectra for atoms in $F=2$. Note that the simulated $E(d)$ does not vary significantly with the exact frequency of the detuning and with the internal state. Right: Relative number of transferred atoms as a function of the 2-photon frequency for different dressing of the $F=1$ state only. We vary in particular the detuning (from left to right): $\vert \Delta_1 \vert = (\Delta_2-9)$ kHz,$(\Delta_2-3)$ kHz,$(\Delta_2+3)$ kHz,$(\Delta_2+10) $ kHz with equal power delivered in the two tones. Sidebands generated by the cloud displacement are visible at frequencies compatible with the trapping along the vertical axis.} \label{fig:sideband}
\end{figure}

In order to study the transfer of the atoms from $\vert 1,-1 \rangle$ to $\vert 2, 1\rangle$, when different external potentials are applied to the two states, we drive a 2 photon transition, as written in section III.B, while applying different two-tone dressings. We perform, in particular, a weak pulse of duration $300$ ms and measure the relative population of the transferred atoms $P_{2}=N_2/(N_1+N_2)$, where $N_1,N_2$ are imaged in TOF by means of a double-state detection technique. Scanning the frequency of the RF photons, we can get the spectrum of the final potential in $F=2$. We repeat the measurements for different powers of the MW-dressing which correspond to different realizations of the two external potentials. In Fig.\ref{fig:spectra} we show the experimental results, for $\Delta_1=-\Delta_2=52$ kHz and $P_{mw-1}=P_{mw-2}=P_{mw}= 10, 10.5, 11, 11.5$ mW, which are fitted with a simple 1D model. We calculate the eigenstates $\psi_i(y)$ and energy spectrum $\xi_i$ of an ideal double-well potential $V(y)=\frac{V_0}{(d/2)^4}(y^2-(\frac{d}{2})^2)^2$ with separation between the wells $d$ and barrier height $V_0$ as adjustable parameters. The wavefunction of the atoms in $F=1$, which we assume to be all in the ground state of the harmonic potential, is simply written as $\phi(y)=\frac{1}{\sqrt{2\pi} \sigma}e^{-(y-y_0)^2/(2\sigma^2)}$. The coupling strength is given by $\Omega(\nu)=A \sum_{i} \frac{C_i}{1+(\nu-\nu_i-\nu_c)^2/w^2}$ with $C_i= \vert \int \psi_i^*(y) \phi(y) dy \vert$ and $\nu_i=\xi_i/(2\pi\hbar)$, where $w$ is the spectral width of the pulse and $\nu_c$ an offset frequency. The parameters $d$, $y_0$ and $\sigma$ have been experimentally determined. This simple model does not account for interactions between the atoms, however it can still provide an understanding of the spectra without the need of too involved calculations. In particular, it reproduces quite well the measurements at higher values of the power $P_{mw}=11.5,11$ mW. In this case the atoms in $F=1$ are mainly pushed on one side along the $y$-axis by the dressed potential $V_{mw-1}$ and thus better couple to just one of the two double-wells (see insets of Fig. \ref{fig:spectra}). For lower $P_{mw}$ the atoms are less displaced from the center of the unperturbed trap, and the initial wavefunction $\phi(y)$ starts to be more symmetrically distributed over the two wells. The worse agreement with the simple model, which is observed for these powers of the dressing, can be explained with the existence of asymmetries on the double-well potential which are not caught by our ideal representation of the potential $V(y)$. We can estimate this asymmetry to be $15-20 \%$ in the range of MW power considered here. This measurement has been performed by fast transferring the atoms to the top of the potential barrier of the double well. The cloud then rapidly fragments in two pieces following a complete oscillation in each of the two wells, from which we can derive the trapping frequencies. By comparison with the simple theory we can, however, address the eigenstates of the external potential which we occupy by transfer the atoms to the $F=2$ hyperfine level. We can also derive an estimation of the barrier height, which is plotted, for comparison, together with the simulations of the potential in Fig.\ref{fig:sideband}. 

Finally we note that the dressing on the $\vert 1,-1\rangle$ state is necessary to maintain a finite wavefunction overlap between the atoms in $F=1$ and $F=2$ along the $z$-axis. Without this correction to the potential of the $F=1$ atoms, the two clouds would be displaced vertically by a distance of $15$ $\mu$m for our experimental parameters, which is larger than their radial extension. A fine tuning of the potentials $V_{mw-1}$ and $V_{mw-2}$ is shown in Fig. \ref{fig:sideband} where transfer spectra have been measured for different detuning $\Delta_1$. A difference in the vertical position within the two clouds $d_z$ leads to the appearance of additional sidebands corresponding to the vertical trapping frequency with coupling elements $\Omega_{F=1, n=1 \rightarrow F=2, n'=2}=\langle F=2,n'=2\vert \Omega_{2ph} e^{\frac{id_z\widehat{p}_z}{\hbar}} \vert F=1,n=1 \rangle$, where $\widehat{p}_z$ is the momentum transfer along $z$-axis. The best wavefunction overlap for the two clouds along the vertical direction is found for $\vert \Delta_1 \vert > \vert \Delta_2 \vert$ with equal MW-power $P_{mw-1}=P_{mw-2}$. 

Finally, by tuning the frequency of the pulse, for a $\pi$ pulse duration $\tau \gg 2\pi/\omega_{ho}$, we can thus address different single external states of the double-well potential in $F=2$. For a given power, the transfer to the ground state, as shown by the spectra, turns out to be significantly less consistent with respect to the high-energy states closer to the barrier height due to bad wavefunction overlap. This problem can be, to some extent, circumvented as the wavefunction of the atoms before the transfer can be independently modified and the wavefunction overlap along the $y$-axis further optimized. 

On the other extreme, fast transfers $\tau \ll 2\pi/\omega_y$ lead to a rather different scenario with the coherent occupation of several vibrational states of the double-well in $F=2$. Once transferred the atoms find themselves out of equilibrium with respect to the new external potential: the single BEC thus splits in two fragments which follow their own dynamics rolling down the potential hill and coming back after an oscillation period to recombine. This configuration presents some similarities with the in-trap splitting of a BEC by means of optical Bragg diffraction \cite{horikoshi2007}, and it could be considered for a matterwave analog of a white light interferometer.
\section{DISCUSSION and CONCLUSIONS}

In the previous paragraphs we have demonstrated the possibility to realize double-well potentials, selectively on the atomic internal state, by the use of micro-wave fields delivered by the chip structures. Furthermore, by applying appropriate Rabi pulses we can transfer the atoms selectively to one external state of the double-well potential. However our peculiar chip device shows some limitations due to the large separation ($\gtrsim 30$ $\mu$m) between the potential minima which is needed to achieve a barrier height safely bigger than the BECs chemical potentials. Given the relatively shallow radial frequencies, we experience a limit in the possibility of adjusting the potential of the state $\vert 1 \rangle$ due to the need of maintaining a sufficient wavefunction overlap along the vertical direction. As a result, the maximum overlap among the wavefunctions in $\vert 1 \rangle$ and $\vert 2 \rangle$ is below $1\%$. This requires long transfer times and non-trivial engineering of the Rabi pulses in order to achieve an useful transfer efficiency. With a different geometry of the chip, with wires size of few microns instead of tens as in our device, or with the different static trapping geometry which we discussed above, a suitable wavefunction overlap can be obtained. 

To conclude, we have proposed a novel beam splitter technique for BECs which does not require a dynamical deformation of the trapping potential. This allows the preparation of an initial clean coherent state in a double-well potential without the formation of spurious, unwanted, excitations. We have numerically shown that the ground state of a double-well potential can be coherently and rapidly populated with excellent fidelity and efficiency without the use of optimal control techniques, with realistic simulation numbers. By studying an experimental case, we present the first realization of a MW-dressed double-well potential selectively on one internal state of $^{87}$Rb and we prove at the same time independent control on the potentials of the two different internal states. We have also characterized the transfer procedure and the addressing of different vibrational states in the double-well. This work shows that to obtain a good transfer efficiency and fidelity in reasonable transfer times, the wavefunction overlap between the initial and final states should be optimized. This is achievable with current state-of-the-art experiments, promoting our technique as a very promising tool in the field of atom interferometry with BECs, e.g. by opening the possibility to monitor the dynamical formation of non-trivial entangled states with metrological gain. We finally note that this system constitutes a potential realization of the long discussed spin-boson model \cite{leggett1987}, which would allow the study of decoherence processes in a controlled environment \cite{scelle2013}.



\section*{Acknowledgements}
We would like to thank Philipp Treutlein, Andrea Trombettoni, Luca Pezz\'e and Wilfried Maineult for fruitful discussions. We acknowledge financial support by the French National Research Agency within the CATS project ANR-09-NANO-039 and by the EU under the project EMRP IND14.

\end{document}